\documentclass[letterpaper,10pt]{article}

\usepackage{opticameet3}
\usepackage{pgfplots}
\usepackage{subcaption}
\DeclareCaptionFont{eightpt}{\fontsize{8}{9.6}\selectfont}
\pgfplotsset{every mark/.append style={solid}}
\usepackage{bm}

\pgfplotsset{compat=1.18}
\usetikzlibrary{fpu}
\usetikzlibrary{svg.path}

\newcommand\authormark[1]{\textsuperscript{#1}}

\usepackage{amsmath,amssymb}
\usepackage[colorlinks=true,bookmarks=false,citecolor=blue,urlcolor=blue]{hyperref}

\definecolor{new_blue}{RGB}{93,169,233}
\definecolor{new_green}{HTML}{0FA3B1}
\definecolor{new_red}{HTML}{DD2D4A}
\definecolor{new_braun}{HTML}{79745C}
\definecolor{new_cherry}{HTML}{880D1E}
\definecolor{new_fildisi}{HTML}{F5EFED}
\definecolor{new_turqouise}{HTML}{508991}
\definecolor{new_purple}{HTML}{473198}
\definecolor{new_gray}{HTML}{67597A}

\definecolor{new_purple_soft}{HTML}{ACA2D2}
\definecolor{new_gray_soft}{HTML}{BCB5C4}
\definecolor{new_red_soft}{HTML}{DE828E}

\definecolor{KITpalegreen}{RGB}{130,190,60}
\definecolor{KITgreen}{rgb}{0,.59,.51}
\definecolor{KITorange}{RGB}{223,155,27}

\newcommand{\vectorrep}[1]{\ensuremath{\underline{#1}}}

\begin{document}

\title{Ultra‑Low‑Rate Information Reconciliation: \\ Repetition Coding or Dedicated Codes?}

\author{Erdem Eray {C}il\authormark{*} and Laurent Schmalen}

\address{Communications Engineering Lab, Karlsruhe Institute of Technology, Hertzstr. 16, 76187 Karlsruhe, Germany\\}

\email{\authormark{*}\texttt{erdem.cil@kit.edu}}

\begin{abstract}
We compare repetition‑based ultra‑low‑rate information reconciliation with dedicated ultra‑low‑rate codes for CV‑QKD. Repetition coding offers a favorable performance–complexity trade‑off, incurring only a moderate error‑rate penalty while reducing decoding complexity by $2\times$, making it attractive for implementation‑constrained systems.
\end{abstract}

\section{Introduction}
In continuous variable-quantum key distribution~(CV-QKD), information reconciliation~(IR) directly limits both secret-key rate and achievable distance~\cite{leverrier08}. At low signal-to-noise ratios~(SNRs), secret-key extraction is only possible if the reconciliation efficiency remains very close to $1$, consequently, high-performance ultra-low-rate error correction is required.

Two competing approaches exist for ultra-low-rate error correction. The first uses dedicated ultra-low-rate low-density parity check~(LDPC) codes~\cite{Gumus2021,PhysRevA.103.062419}, which can achieve a good frame error rate~(FER) performance but typically require long block lengths and many decoding iterations; for some designs, decoding complexity increases further as the rate decreases~\cite{Gumus2021}. The second uses repetition-based IR~\cite[Sec.~5.2]{leverrier09}, which maps a higher-rate mother code to lower effective rates through repeated frames while naturally enabling rate adaptivity.

The practical question is not only \emph{which approach gives the best FER}, but \emph{which approach gives the best FER-complexity operating point} under implementation constraints (decoder throughput, latency and hardware cost).

In this paper, we explain the repetition coding within the multidimensional reconciliation~(MDR)~\cite{leverrier08} framework and compare repetition-based and dedicated ultra-low-rate designs at the same target rate. Our results show that dedicated codes provide the best FER, while repetition-based schemes can substantially reduce decoding complexity with only a moderate FER penalty.

\section{Rate Extension by Repetition Coding}
Throughout this paper, underlined symbols represent column vectors, and $\odot$ represents the Hadamard product.

\subsection{Protocol}
We assume reverse reconciliation with MDR~\cite{leverrier08}. For the first frame, Alice sends quantum states $\vectorrep{x}^{(1)}$, and Bob measures $\vectorrep{y}^{(1)}$. Bob draws a binary raw key $\vectorrep{u}^{(1)}$, computes the LDPC syndrome $\vectorrep{s}^{(1)}=\mathbf{H}\vectorrep{u}^{(1)}$ using the parity-check matrix $\mathbf{H}$ of an LDPC code with rate $R_\mathrm{LDPC}$ and derives MDR rotations $\vectorrep{\phi}^{(1)}$ from $\vectorrep{y}^{(1)}$ and $\vectorrep{u}^{(1)}$. He then transmits $\vectorrep{s}^{(1)}$ and $\vectorrep{\phi}^{(1)}$ over the classical channel.

Each subsequent frame replicates the first through repetition coding. For the frame $\ell\in\{2,\dots,N_{\mathrm{rep}}\}$, Bob draws an independent key $\vectorrep{u}^{(\ell)}$,  binary summation of the keys \mbox{$\vectorrep{s}^{(\ell)}=\vectorrep{u}^{(\ell)} \oplus \vectorrep{u}^{(1)}$} and transmits $\vectorrep{s}^{(\ell)}$ together with rotations $\vectorrep{\phi}^{(\ell)}$ computed from fresh measurements $\vectorrep{y}^{(\ell)}$ and $\vectorrep{u}^{(\ell)}$. This yields the repetition rate $R_\mathrm{rep}=1/N_{\mathrm{rep}}$.

To decode, Alice firstly computes the log-likelihood ratio of the symbols in the $i$th frame $\vectorrep{L}^{(i)}$ for \mbox{$i \in \{1,\dots,N_{\mathrm{rep}}\}$}, and decoding starts from
\begin{equation}
 \vectorrep{L}_\mathrm{init} = \vectorrep{L}^{(1)} + \sum_{\ell=2}^{N_{\mathrm{rep}}} (-2\vectorrep{s}^{(\ell)}+1) \odot\vectorrep{L}^{(\ell)} .   \label{eqn:repetition}
\end{equation}
Repetition effectively creates a virtual channel seen by the higher-rate decoder. The reconciliation efficiency $\beta$ can be written as
\begin{equation*}
    \beta = \frac{R}{C} = \frac{R_\mathrm{rep}R_\mathrm{LDPC}}{C} = \beta_\mathrm{rep}\beta_\mathrm{LDPC},
\end{equation*}
where $\beta_\mathrm{LDPC}$ is the reconciliation efficiency of the LDPC code. For an SNR $x$, the channel capacity is \mbox{$C=\frac{1}{2}\log_2(1+x)$}, and \mbox{$\beta_\mathrm{rep}=\log_2(1+N_\mathrm{rep}x)/(N_\mathrm{rep}\log_2(1+x))$} captures the asymptotic efficiency loss due to repetition~\cite[Sec.~5.2]{leverrier09}. Fig.~\ref{fig:beta_rep} illustrates $\beta_\mathrm{rep}$ for the representative case \mbox{$R=0.005$}.

To analyze the protocol's security, we obtain an equivalent compact formulation by stacking the frames into
\mbox{$\tilde{\vectorrep{a}}=( (\vectorrep{a}^{(1)})^T \cdots (\vectorrep{a}^{(N_\mathrm{rep})})^T)^T$}.
Then Alice sends $\tilde{\vectorrep{x}}$, Bob measures $\tilde{\vectorrep{y}}$, and Bob computes
$\tilde{\vectorrep{s}}=\tilde{\mathbf{H}}\tilde{\vectorrep{u}}$ with
\begin{equation*}
\tilde{\mathbf{H}}=\begin{bmatrix}
{\mathbf{H}} & \mathbf{0} \\
{\mathbb{I}} & \mathbb{I}
\end{bmatrix}.
\end{equation*}

Here, $\mathbb{I}$ and $\mathbf{0}$ denote the identity and zero matrices, and the resulting rate is $R=R_\mathrm{rep}R_\mathrm{LDPC}$. Bob also transmits MDR rotations $\tilde{\vectorrep{\phi}}$ derived from $\tilde{\vectorrep{y}}$ and $\tilde{\vectorrep{u}}$. Under collective attacks, $\tilde{\vectorrep{x}}$ and $\tilde{\vectorrep{y}}$ are i.i.d., and $\tilde{\vectorrep{u}}$ is uniform i.i.d. and independent of $\tilde{\vectorrep{y}}$; therefore, the security proof of~\cite{leverrier08} remains applicable.

\subsection{Decoding Complexity}
We estimate the complexity of iterative belief-propagation decoding by counting message updates in the Tanner graph. On a per-information-bit basis, this scales with (average variable node~(VN) degree) $\times$ (average number of iterations) for the code constructions we use~\cite{5407615}. Because degree-1 VNs do not participate in message updates, they are excluded when computing the average VN degree. In repetition-based decoding, the preprocessing in (\ref{eqn:repetition}) adds negligible overhead compared with iterative decoding, so it is omitted from the complexity metric.

\section{Simulation Results}
\begin{figure}[t!]
  \centering
  \includegraphics[width=\linewidth]{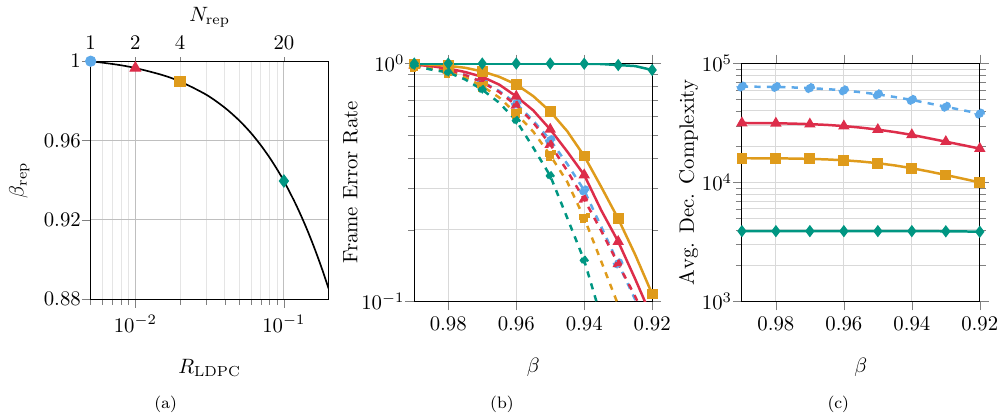}%
  \centering
  \setbox0=\hbox{%
    \begin{tikzpicture}
      \begin{axis}[hide axis, width=4cm, height=4cm,
        legend to name=figtwolegend,
        legend columns=4,
        legend style={/tikz/every even column/.append style={column sep=0.45em}, align=center, draw=black, fill=white, font=\normalsize,
        /tikz/every odd column/.append style={column sep=0.4em},
        /tikz/every even column/.append style={column sep=1.2em},
        },
        legend image post style={only marks},
        legend cell align={left},
      ]
        \addplot[new_blue, line width=1.2pt, mark=*, dashed] coordinates {};
        \addlegendentry{$R_\mathrm{LDPC}=0.005$};
        \addplot[new_red, line width=1.2pt, mark=triangle*] coordinates {};
        \addlegendentry{$R_\mathrm{LDPC}=0.01$};
        \addplot[KITorange, line width=1.2pt, mark=square*] coordinates {};
        \addlegendentry{$R_\mathrm{LDPC}=0.02$};
        \addplot[KITgreen, line width=1.2pt, mark=diamond*] coordinates {};
        \addlegendentry{$R_\mathrm{LDPC}=0.1$};
      \end{axis}
    \end{tikzpicture}%
  }%
  \pgfplotslegendfromname{figtwolegend}
  \caption{Repetition efficiency $\beta_\mathrm{rep}$, frame error rate and average decoding complexity for LDPC mother-code rates $R_\mathrm{LDPC}$ and reconciliation efficiency $\beta$. In subfigures (b) and (c), dashed curves correspond to no repetition ($N_\mathrm{rep}=1$), while solid curves correspond to the repetition factors indicated in subfigure (a) to obtain an overall rate of $R=0.005$.}
  \label{fig:FER_Complexity_plot}
  \phantomsubcaption\label{fig:beta_rep}%
  \phantomsubcaption\label{fig:FER}%
  \phantomsubcaption\label{fig:complexity}%
\end{figure}
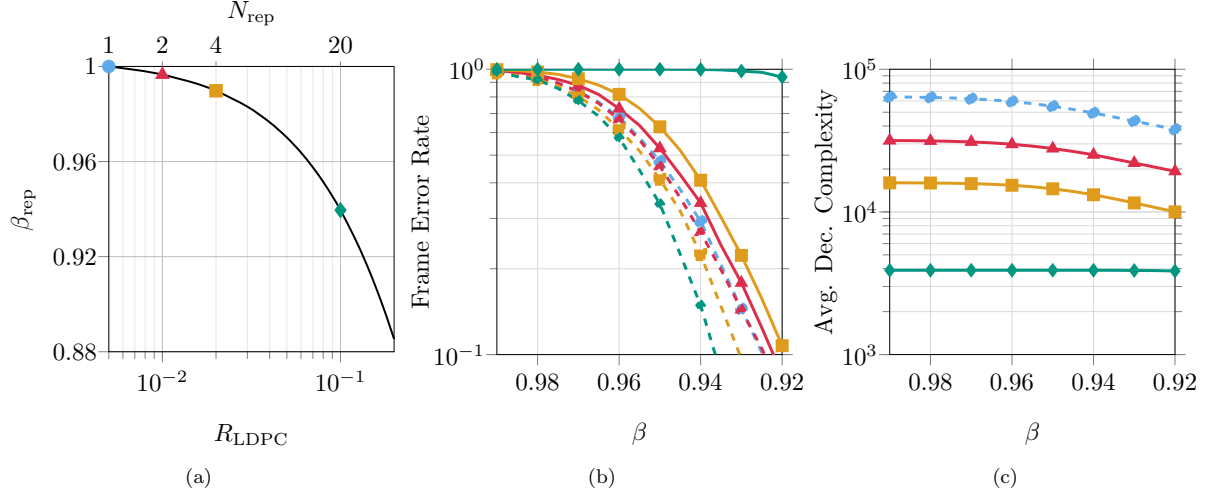

We target an overall rate of $R=0.005$. Following~\cite{Gumus2021}, we optimize protographs for $R_\mathrm{LDPC}=0.005,0.01,0.02,0.1$ (markers in Fig.~\ref{fig:beta_rep}). The lifted block length is chosen such that, after repetition, the effective block length is approximately $2\cdot 10^6$. We assume the four-state protocol~\mbox{\cite[Sec.~5.5]{leverrier09}} and use flooding-schedule sum-product decoding up to 200 iterations.

Figure~\ref{fig:FER} shows the FER curves for the unrepeated designs ($N_\mathrm{rep}=1$, dashed lines). We then apply repetition factors $N_\mathrm{rep}=2,4,20$ to the $R_\mathrm{LDPC}=0.01,0.02,0.1$ codes, respectively, to obtain the same overall rate $R=0.005$ (solid lines). As expected, the FER degrades with a larger number of repetitions; the penalty is most pronounced for $R_\mathrm{LDPC}=0.1$ and decreases for lower mother-code rates, consistent with the $\beta_\mathrm{rep}$ behavior in Fig.~\ref{fig:beta_rep}.

Overall, the dedicated $R_\mathrm{LDPC}=0.005$ code achieves the best FER. However, repetition coding on \mbox{$R_\mathrm{LDPC}=0.01$} or $0.02$ incurs only a moderate FER penalty while reducing decoding complexity by about $2\times$ and $4\times$, respectively (Fig.~\ref{fig:complexity}). Therefore, in the high-FER operating region relevant to long-distance CV-QKD, repetition-based reconciliation offers a compelling implementation trade-off.

\section{Conclusion}
Dedicated ultra-low-rate codes provide the best FER, but repetition-based schemes offer a stronger practical FER-complexity trade-off. When decoder complexity, throughput, or latency is the dominant constraint, repetition-based IR is the more compelling option.

\bibliographystyle{opticajnl}
\footnotesize{\bibliography{sample}}

\end{document}